\documentclass[sn-basic,iicol]{sn-jnl}

\usepackage{caption}
\usepackage{subcaption}
\usepackage{gensymb}
\usepackage{amsmath}

\usepackage{soul}
\usepackage{color}
\usepackage{placeins}
\usepackage{float}
\usepackage{xcolor}

\jyear{2022}

\begin{document}

\title[X-ray PTV With Advanced Tracers]{Enhanced Laboratory X-ray Particle Tracking Velocimetry With Newly Developed Tungsten-Coated $O$(50~$\mu$m) Tracers}


\author*[1]{\fnm{Jason T.} \sur{Parker}}\email{jtparker@berkeley.edu}
\author[2]{\fnm{Jessica} \sur{DeBerardinis}}\email{jessica.deberardinis@ultramet.com}
\author[1]{\fnm{Simo A.} \sur{Mäkiharju}}\email{makiharju@berkeley.edu}

\affil*[1]{\orgdiv{Mechanical Engineering}, \orgname{UC Berkeley}, \orgaddress{\street{Hearst Ave.}, \city{Berkeley}, \postcode{94720}, \state{CA}, \country{USA}}}
\affil[2]{\orgname{Ultramet}, \orgaddress{\street{Montague St.}, \city{Pacoima}, \postcode{91331}, \state{CA}, \country{USA}}}


\abstract{
Tracer particles designed specifically for X-ray particle tracking and imaging velocimetry (XPTV and XPIV) are necessary to widen the range of flows that can be studied with these techniques. In this study, we demonstrate in-lab XPTV using new, custom-designed $O$(50~$\mu$m) diameter tungsten-coated hollow carbon spheres and a single energy threshold photon counting detector. To explore the measurement quality enhancement enabled by the new tracer particles and photon counting detector, a well understood Poiseulle pipe flow is measured. The data show agreement with the analytical solution for the depth-averaged velocity profile. The experiment also shows that the tungsten-coated particles achieve higher contrast and are better localized than previously available silver-coated particles, making faster and more precise measurements attainable. The particles are manufactured with a readily scalable chemical vapor deposition process.

We further show that laboratory XPTV is practical with currently available energy-resolving photon counting detectors (PCDs), despite their presently lower spatiotemporal resolution compared to scintillating detectors. This finding suggests that energy-thresholding identification of different classes of tracers is feasible, further motivating the exploration of the X-ray tracer particle design space. The latest generation of PCDs are incorporating multiple energy thresholds, and have higher count rate limits. In the near future one could potentially expand on the work presented and track multiple tracer species and scalar fields simultaneously.
}


\maketitle

\section{Introduction}\label{sec:intro}

Particle image velocimetry (PIV) and particle tracking velocimetry (PTV) at visible wavelengths are quantitative flow visualization techniques capable of resolving 2D and 3D flow fields that are readily compared to theory and simulation (\cite{m_raffel_particle_2018}). With no probe to disturb the flow field, these techniques are favored throughout the experimental fluid dynamics community.

In a typical 2D PIV or PTV measurement of a flow, tracer particles are illuminated with a laser sheet. Images of the tracer particle motion in the fluid are captured by a high-speed camera. The tracer particles are ideally small enough and density matched with fluid such that they follow the flow accurately without influencing the flow itself. A primary limitation of PIV and PTV is that they require optical access for the laser light to reach the tracer particles, and for the scattered light to reach the camera sensor. For many flows, either due to the opacity of the fluid itself or the surrounding material, optical access is impossible at visible wavelengths. Metal foam flows and natural soil flows, for example, must take place in opaque containers and surroundings.  Solid-liquid flows such as blood and silty water can also be opaque to visible light. Even in multiphase flows where both phases are individually transparent – such as air-water flows – refraction across multiple moving, deforming interfaces makes the flow effectively opaque. Standard flow visualization techniques struggle to yield data on these flows, which together cover substantial swaths of fluid dynamics.

Photons in the X-ray range, however, can pass through many materials that are opaque in the visible light wavelengths, and they have a refractive index near unity. As a result, XPIV and XPTV can overcome the optical access limitations of PIV and PTV. If XPIV and XPTV can be developed into practical techniques with laboratory equipment, the impact would be significant. Three decades of PIV and PTV algorithm development could be applied to previously inaccessible systems such as biological flows (\cite{park_x-ray_2016, antoine_flow_2013, kim_x-ray_2006, jamison_x-ray_2012, krebs_initial_2020}), multiphase flows (\cite{ganesh_bubbly_2016, makiharju_time-resolved_2013, ganesh_bubbly_2016, makiharju_dynamics_2017, yoon_image_2018}), and internal flows (\cite{lappan_x-ray_2020, liu_lagrangian_2021}). In fact, multiphase flow studies already occasionally use other X-ray techniques for measuring the phase fraction.

\cite{lee_x-ray_2003} demonstrated XPIV for the first time at a synchrotron by measuring the 2D-projected flow profile in a round pipe, and this work provides a useful point of comparison for this study. Recently \cite{ge_aps_2021}, used XPIV at a synchrotron to investigate a cavitating shedding flow, the kind of study that would benefit if it could be conducted in a laboratory. In general, research to extend XPIV and XPTV (\cite{dubsky_computed_2009, fouras_three-dimensional_2007}) has focused on work at synchrotrons. However, limiting XPIV and XPTV to synchrotrons hinders the pace of research by imposing practical limitations such as beam time, cost, and location. Furthermore, synchrotrons typically have illuminated areas on the order of a few millimeters, which constrains the experiment geometry and accessible parameter range. For example, \cite{ge_aps_2021} could examine a larger domain with the larger illuminated area of a laboratory source. Enabling XPTV at the laboratory-scale is imperative for it to become a more widely useful fluid dynamics measurement technique.

Although some progress has been made recently towards practical in-lab X-ray PTV (\cite{makiharju_tomographic_2022, parker_experimentally_2022, bultreys_x-ray_2022}), the technique's applicability remains limited. Formidable obstacles exist to obtaining a sufficient signal-to-noise ratio (SNR) given the dimness of laboratory X-ray sources compared to synchrotrons, and the difficulty of finding high-contrast particles that are also good flow tracers. \cite{poelma_measurement_2020} reviewed various techniques for measuring multiphase flows and also discussed the challenges of performing laboratory XPIV and XPTV. For brevity, throughout the rest of this paper, the discussion will focus on XPTV, although many of the findings apply to XPIV as well.

In order to achieve particle image contrast, most previous in-lab XPTV experiments have relied on particles that are either near order of magnitude density mismatched with the fluid, are on the order of a millimeter in diameter, or both (\cite{lappan_x-ray_2020, lee_development_2009, heindel_x-ray_2008}). However, most particles that are not neutrally buoyant bias the velocity measurements and become unevenly distributed. Large particles, even if they are nominally density matched, limit the spatial resolution and may not trace the flow, particularly in large velocity gradients or if flow features are on the order of magnitude of the particle size. Recently, \cite{makiharju_tomographic_2022} investigated a creeping flow with 60~$\mu$m tracer particles. In that study, we noted the significant limitations imposed by the slow scan times that were necessary to achieve particle image contrast. As part of that study, we identified improving the tracer particles to be of paramount importance to laboratory XPTV.

XPTV imposes unique requirements on tracer particles that conflict with one another. XPTV tracer particles must attenuate X-rays significantly more (or less) than the ambient fluid to generate contrast, which favors large, density-mismatched particles. At the same time, tracer particles must be neutrally buoyant and have sufficiently low inertia to trace the flow accurately, which favors small, density-matched particles. In addition, small tracer particles maximize the spatial resolution of the measurement. \cite{parker_experimentally_2022} developed and validated methods for simulating XPTV images to predict the performance of existing and conceptual XPTV tracer particles before investing in expensive custom tracer particle manufacturing. These tools allow us to systematically balance the design criteria for XPTV tracer particles. Thanks to modern manufacturing processes it is now feasible to develop $O$(10~$\mu$m) prototype custom tracer particles for XPTV applications.

In this paper we introduce new hollow carbon (C) tungsten-coated (W) microsphere XPTV tracer particles (CW) that were designed using the simulation tools in \cite{parker_experimentally_2022}. We compare the performance of the CW tracer particles to the performance of silver-coated hollow glass tracer particles (AGSF-33), a previously available visible light PTV tracer particle. Tracer particle performance is evaluated by measuring Poiseulle pipe flow with 2D-projected XPTV. Additioanlly, images of the tracer particles are captured with a variable single-energy-threshold photon counting detector (PCD), demonstrating that PCDs can be used for XPTV in the laboratory. PCDs can attain a higher SNR than scintillating detectors. Using PCDs also opens up the possibility of future experiments tracing multiple particle species, scalar fields, or both with K-edge material detection.

The paper is organized as follows: in section 2 we discuss the methodology and materials, section 3 discusses the properties of the CW and AGSF-33 tracer particles, section 4 presents and discusses the results followed by the conclusions in section 5.

\section{Methodology}\label{sec:methd}

\subsection{Experimental Setup}\label{ssec:expset}
\begin{figure}
    \centering
    \includegraphics[width=\columnwidth]{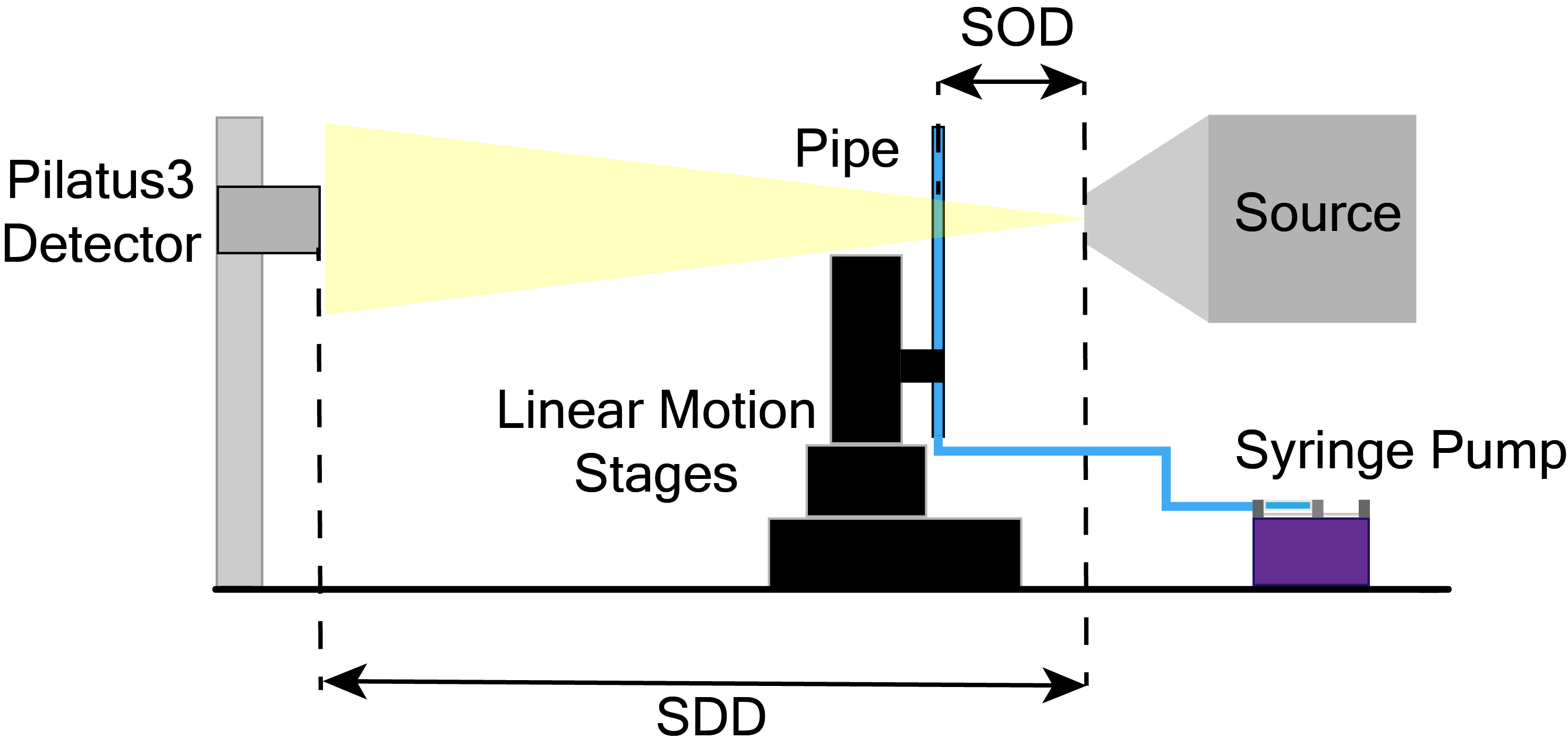}
    \caption{The experimental setup (not to scale). A syringe pump pushes glycerol and tracer particles through a vertically mounted pipe.}
    \label{fig:expsetup}
\end{figure}

The experimental setup can be seen in figure \ref{fig:expsetup}. The source to detector distance (SDD) and source to object distance (SOD) are 500~mm and 38~mm, respectively. The SOD is measured from the X-ray emitting tungsten target surface to the center of the pipe. The source emits a conical beam, which geometrically magnifies objects in the field-of-view (FOV) by a factor of approximately $M=13.1$. The detector has a panel size of 83.8~mm $\times$ 33.5~mm. Due to geometric magnification, the FOV in the object plane is 6.37~mm $\times$ 2.55~mm at the central plane of the pipe.

The available quantity of the new custom particles made a closed loop impractical. The tracer particles are seeded directly into the 6.35~mm (0.25~in.) inner diameter pipe by pouring a particle-glycerol mixture into the top of the pipe at an angle, leaving room for the air in the pipe to exit. The mixture is at equilibrium at 23.7~$\degree$C as measured by an Ambient Weather WH31E temperature and humidity sensor adjacent to the X-ray source. A Harvard Apparatus 11 Plus syringe pump pushes the glycerol at a constant flow rate $Q = 0.215$~mL/min. The 11 Plus syringe pump has a 0.9$\degree$ step rotation for high accuracy flow rates. The Reynolds number of the flow, $Re = Ud/\nu = 0.0018$, is of comparable order of magnitude as Lee and Kim's experiment at a synchrotron, which, based on data they provide, is calculated to be $Re_{LK} = 0.0052$. Here, $U$ is the centerline velocity and $d$ is the pipe inner diameter. The laminar development length for the current experiment is $l = 0.06 Re D = 0.69~\mu$m. Given that the pipe length upstream of the measurement FOV is significantly longer than the entrance length, the flow can be considered fully developed. 

The pipe is mounted on a 3-axis linear motion system comprised of three Thor Labs NRT150 stages. The NRT150 stages have an on-axis accuracy of 2~$\mu$m and a minimum incremental motion of 0.1~$\mu$m. A Starrett No.98 machinist level with a sensitivity of 0.024$\degree$ is used to ensure that the pipe is mounted vertically.

The Dectris Pilatus3 PCD and YXLON FXE225.99 X-ray source are the same as used in our prior study (\cite{parker_experimentally_2022}). The detector has a 1~mm thick CdTe panel to capture photons in pixels of size 172 $\times$ 172 $\mu$m. Sixteen tiles comprise the detector panel, leading to a 487 $\times$ 195 pixel detection area. The Dectris Pilatus3 can achieve frame rates up to 500~Hz. Due to the relatively dim X-ray source use, we chose a low flow speed such that images captured with 15~ms exposure times would not exhibit motion blur, but would have sufficient photon counts to achieve a usable SNR. Frame rates up to about 70~Hz could be used based on this exposure time. For PTV, however, it is preferable to have a notable particle shift between each frame, so we captured images at 2.5~Hz to allow sufficient particle motion between frames. The detector energy threshold is set to 20~keV to reduce photon scattering noise. The X-ray source is operated at an acceleration voltage of 55~kV with a target current of 500~$\mu$A.

\subsection{Image Processing Methods}\label{ssec:improc}
\subsubsection{Filtering and PIV Software}
Raw particle images are corrected with a flat field correction map, inverted, and then imported to LaVision DaVis 8.4 for filtering and PTV processing. We use standard DaVis algorithms without modification to demonstrate that XPTV data can be processed with existing algorithms. Details on the filters and particle tracking settings can be found in appendix \ref{app:imgproc}.

\subsubsection{Scaling and Uncertainty}
Image pixel-to-mm scaling is done by identifying the pipe inner diameter in the image. The pipe inner diameter is known to be 6.35~mm, and we count the number of pixels required to traverse the inner diameter of the pipe. The time step $\Delta$t is the time between the beginning of each exposure. For all of the images collected in this study, $\Delta$t = 400~ms to ensure that the particles travel sufficiently far between exposures for particle displacement to be significantly larger than the uncertainty in localizing the particle centroid. Given the 15~ms exposure time, the acquisition duty cycle for these experiments is 3.75\%.

As noted in section \ref{sec:intro} and \cite{makiharju_tomographic_2022}, the density mismatch between fluid and particles can be a significant source of uncertainty for XPTV and XPIV. Laboratory systems with dim X-ray sources are currently limited to slow flows where the density mismatch has a more pronounced effect. Discussion of the effects of buoyancy and flow tracing accuracy is reserved for section \ref{sec:partprop}.

There are additional uncertainty contributions unique to XPIV and XPTV, though. Most laboratory X-ray sources have conical beams, leading to geometric magnification, as mentioned previously. The geometric magnification factor $M$ is a function of the distance from the source, so it changes along the beam direction. Accordingly, the FOV will change over the depth of the experiment geometry. In our experiment, the FOV changes from approximately 5.84~mm $\times$ 2.33~mm at the front of the pipe to 6.90~mm $\times$ 2.76~mm at the back. In all planes along the beam direction, the entire pipe is in the FOV. Similarly, the motion of a particle image is a function of the magnification. Images of particles closer to the source will appear larger and will move faster than images of particles farther from the source, even when the particles are the same size and have same exact velocity.

\begin{figure}
    \centering
    \includegraphics[width=0.6\columnwidth]{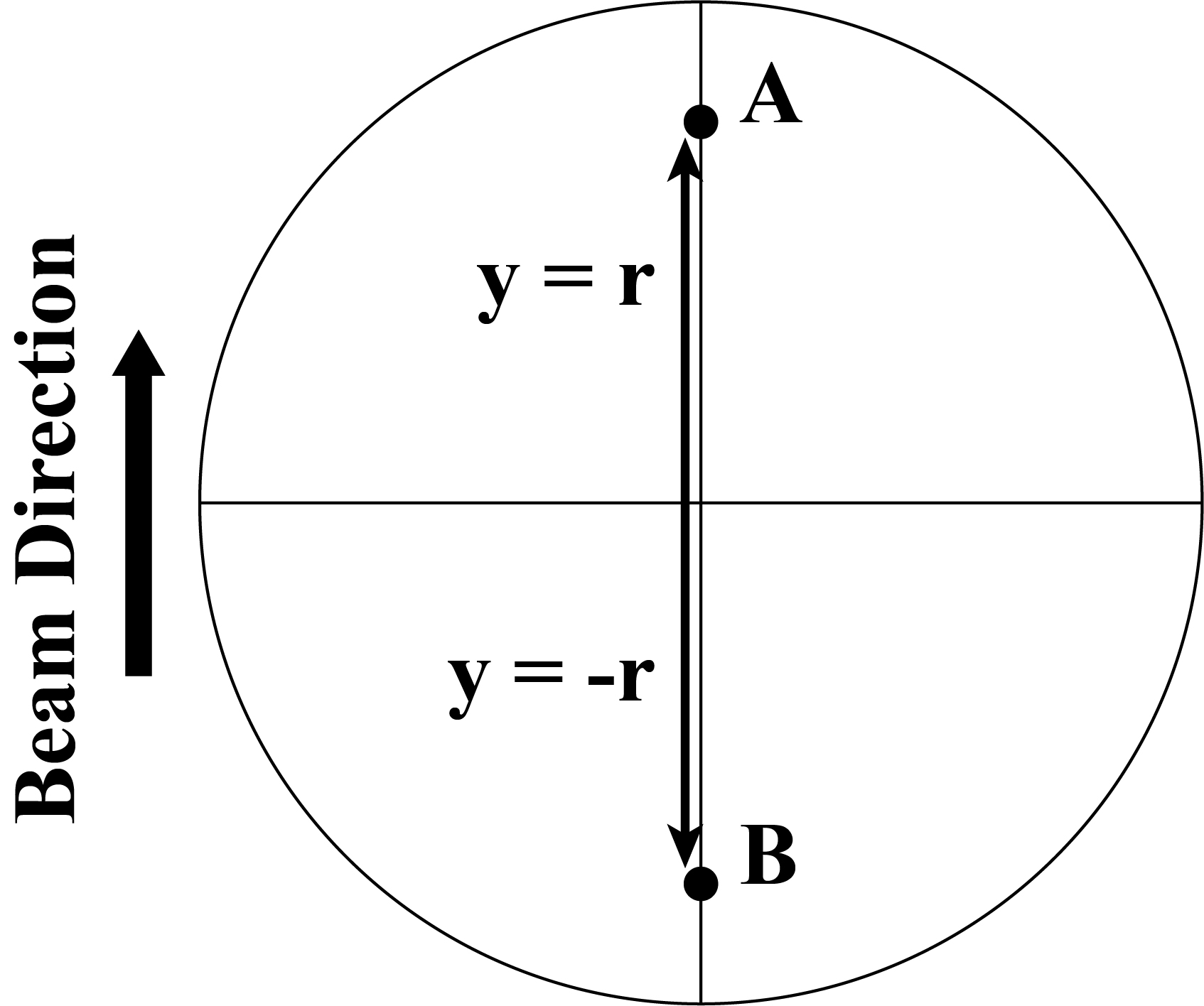}
    \caption{Particle A will be magnified less than particle B because it is farther from the X-ray source. Particle A will also appear to move slower than particle B.}
    \label{fig:partlocs}
\end{figure}

Consider, for example, two identical particles in a Poiseulle pipe flow, each at the same radius, but one particle is closer to the source than the other, as shown in figure \ref{fig:partlocs}. Being at the same radius, both particles will travel at the same velocity. Since particle A is farther from the X-ray source, it will be magnified less than particle B, and will appear to move less than particle B even though it has traveled the same real distance for any given time interval. Given that there is no information on particle depth in a 2D-projected measurement, the pixel-to-mm scaling assumes that all particles are located in the central plane at $y=0$. Equation \ref{eq:motion} gives the measured displacement $\Delta z$ of a particle that has moved a real distance $\Delta Z$. Recall that SOD is measured to the center of the pipe.
\begin{equation} \label{eq:motion}
    \Delta z = \frac{M(y)}{M(y=0)} \Delta Z = \frac{SOD \cdot \Delta Z}{SOD + y}
\end{equation}
The relative error, $\epsilon$, in the measured displacement versus the real displacement will increase with particle distance $y$ from central plane and is given by
\begin{equation} \label{eq:magerror}
    \epsilon = \frac{ \lvert \Delta z - \Delta Z \rvert }{ \Delta Z} 
    = 1 - \left(1 + \frac{\lvert y \rvert}{SOD} \right)^{-1}.
\end{equation}
The velocity relative error is the same as equation \ref{eq:magerror}. Note that the relative error depends on SOD and domain size, but not directly on SDD. We denote the ratio of the experiment depth $\delta = 2y_{max}$ to the SOD as the magnification aspect ratio. A good rule of thumb to avoid magnification error is to keep the magnification aspect ratio below 0.2. The maximum relative magnification error would then be 10\%. In practice, SDD could affect the error by modifying how many pixels a particle image occupies on detector, contributing to the resolution error of localizing a particle centroid.

For this experiment, where the SOD is 38~mm and the maximum $y=3.2$~mm, the relative magnification error never exceeds 9\%, which occurs at the pipe wall as the velocity approaches zero. At the center of the pipe, where the calibration is defined, the relative magnification error is zero. Experimenters with large depths of field should be cautious about the error introduced by magnification effects. Using fully tomographic XPTV, such as in \cite{makiharju_tomographic_2022}, avoids issues with geometric magnification altogether.  Alternatively, if the tracer particles are sufficiently monodisperse, one could determine the depth location based on the particle image size, correct for the magnification error, or both.

\subsection{Depth-Averaged Velocity Profile}
\begin{figure}
    \centering
    \includegraphics[width=0.7\columnwidth]{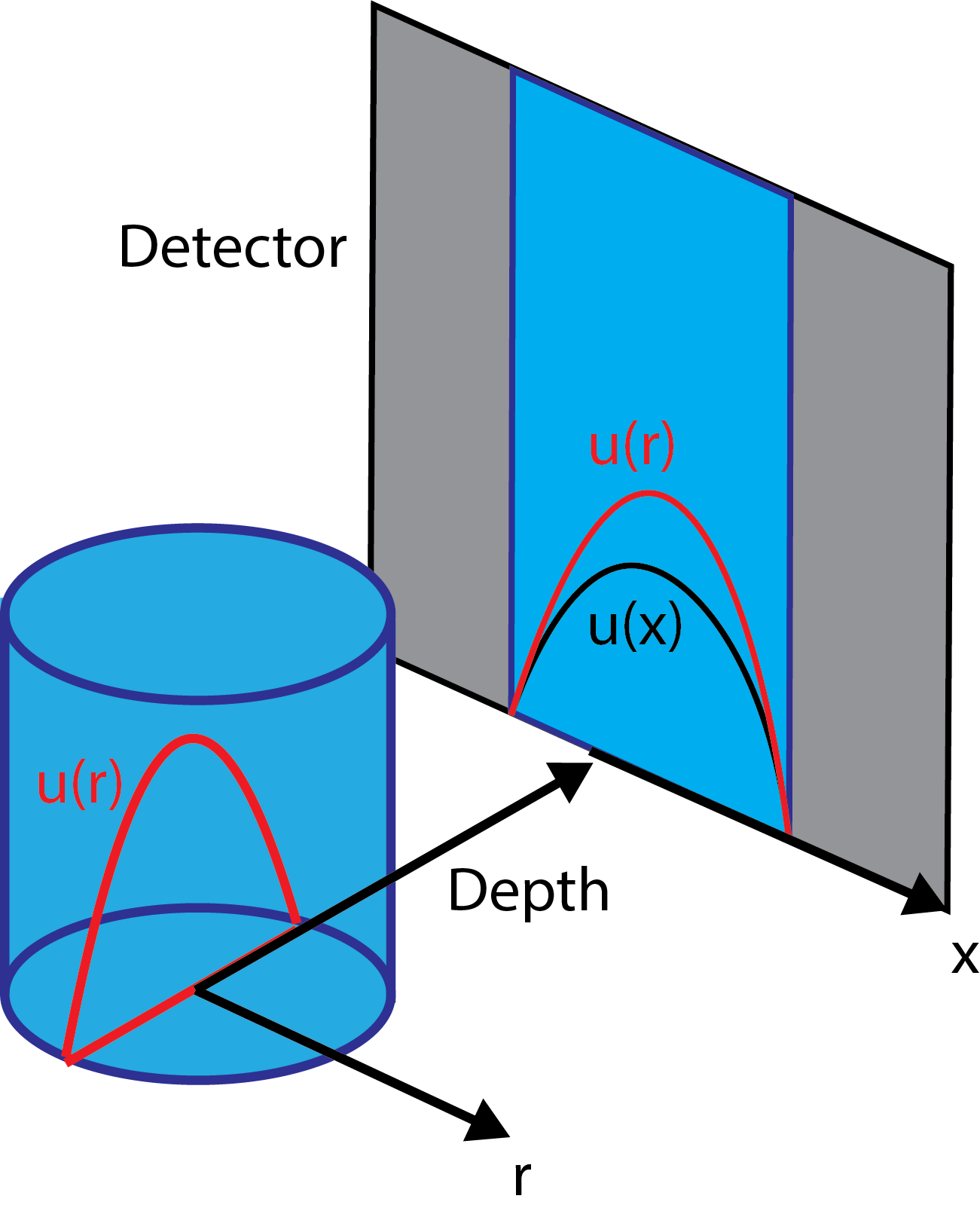}
    \caption{The depth-wise direction is towards the detector. The analytical velocity profile is depth-averaged.}
    \label{fig:davp}
\end{figure}

In this study, 2D-projected images of flow in a pipe are captured. Particles all along the X-ray beam direction (depth-wise direction $y$ in figure \ref{fig:partlocs}) of the pipe are imaged at the same $x$-location. As a result, for particle tracking velocimetry, one is not sampling a single value but rather the distribution of velocities along the depth-wise direction of the pipe. \cite{lee_x-ray_2003} compared the measured average velocity of the depth-wise distribution to the analytical depth-averaged velocity profile (DAVP). Figure \ref{fig:davp} depicts the depth-wise velocity variation in the pipe. \cite{lee_x-ray_2003} assumed a uniform distribution of ideal tracer particles throughout the pipe, and that all particles contribute to the distribution equally. Under these conditions, the DAVP for a parallel-beam projection is simply the depth integrated average velocity of the fluid, given by
\begin{equation} \label{eq:davp}
\begin{split}
\left<u(x,y)\right>_y &= \frac{4 Q}{3 \pi R^2} \left[1 - \left(\frac{x}{R}\right)^2\right]
= \frac{2}{3} u(r=x)
\end{split}
\end{equation}
where $\left<u(x,y)\right>_y$ is the DAVP, $u(r)$ is the analytical solution for Poiseulle pipe flow, $R$ is the pipe radius, and $Q$ is the volumetric flow rate. Hidden in equation \ref{eq:davp} is a cosine of the angle between the radius and the $x$-direction on the detector. Since this cosine is unity for this axisymmetric system with vertical alignment, it is omitted from equation \ref{eq:davp} for clarity. 
For consistency throughout the paper, the DAVP will be written as a function of $x$, while the cross-sectional velocity profile will be written as a function of radius $r$. The varying magnification due to a cone-beam instead of a parallel beam is also not considered here since the magnification error is relatively small. 

It is worth noting that Lee and Kim compared their PIV measurement to the DAVP, whereas we are comparing a PTV measurement to the DAVP. \cite{fouras_three-dimensional_2007} points out that PIV calculates a modal average velocity, not the mean average that Lee and Kim compared against. By comparing a PTV DAVP measurement to the analytical DAVP, we omit any error associated with the discrepancy between the distribution mode and mean.

\section{Tracer Particle Properties}\label{sec:partprop}

\subsection{Silver-Coated Hollow Glass Spheres (AGSF-33)}\label{ssec:agsf}
Silver-coated hollow glass spheres are commonly used in visible wavelength PTV applications, so they are readily available for use in XPTV. The silver coating attenuates X-rays better than uncoated plastic or glass microspheres owing to the relatively high mass attenuation of silver. Higher mass attenuation elements such gold, silver, and tungsten achieve better image contrast provided that the surrounding materials are comprised of lower mass attenuation elements such as hydrogen, oxygen, and carbon. The opposite is true for dense, heavy element fluids like liquid metals. In these experiments, we use AGSF-33 silver-coated hollow glass spheres, a previously available off-the-shelf tracer particle. AGSF-33 particles were previously shown to be suitable XPTV tracer particles (\cite{parker_experimentally_2022, makiharju_tomographic_2022}).

For these experiments, the AGSF-33 particles are sieved to between 45 and 53~$\mu$m to improve monodispersity. We assume a uniform distribution across this size range. Although the AGSF-33 particles are on average matched with the density of water, they, like most tracer particles, have a non-uniform density within a batch. According to the manufacturer, Potters Beads, the particles have density between 0.9 and 1.1~g/cm$^3$. Density mismatches with the fluid induce relative motion due to buoyancy, which can bias velocity measurements, particularly in low speed flows if the velocity due to buoyancy is the same order of magnitude as the flow itself. Although buoyancy-induced velocity is often neglected in visible light PTV, early XPTV experiments will study low speed flows where buoyancy-induced velocity should be considered. When the relative velocity is low enough that $Re(u_{rel}) \ll 1$, and Stokes drag is appropriate, the terminal speed with which particles will rise or settle in a fluid can be calculated from equation \ref{eq:stokesDrag}.
\begin{equation} \label{eq:stokesDrag}
    u_{St} = d_p^2 \frac{\rho_{fluid} - \rho_{particle}}{18 \mu} g
\end{equation}
Here, $d_p$ is the particle diameter, $\rho_i$ is density of $i$, $\mu$ is the dynamic viscosity of the fluid, and $g$ is the acceleration due to gravity. The Stokes settling speed can be thought of as a bias error in the velocity measurement. The Stokes settling speed is proportional to the particle diameter squared, which incentivizes the use of smaller particles. Smaller particles also have a lower Stokes number, which is discussed in section \ref{ssec:particleerr}. The trade off of using a smaller particle, however, is less X-ray attenuation, which, according to the Beer-Lambert law, is exponentially dependent on the particle size. These opposing design criteria are an example of the challenges and opportunities of the XPTV tracer design space. Exploring smaller tracer particle designs that incorporate composite- and single-element hollow shells of high mass attenuation elements while remaining neutrally buoyant is crucial for improving XPTV.

\begin{figure}
    \centering
    \includegraphics[width=\columnwidth]{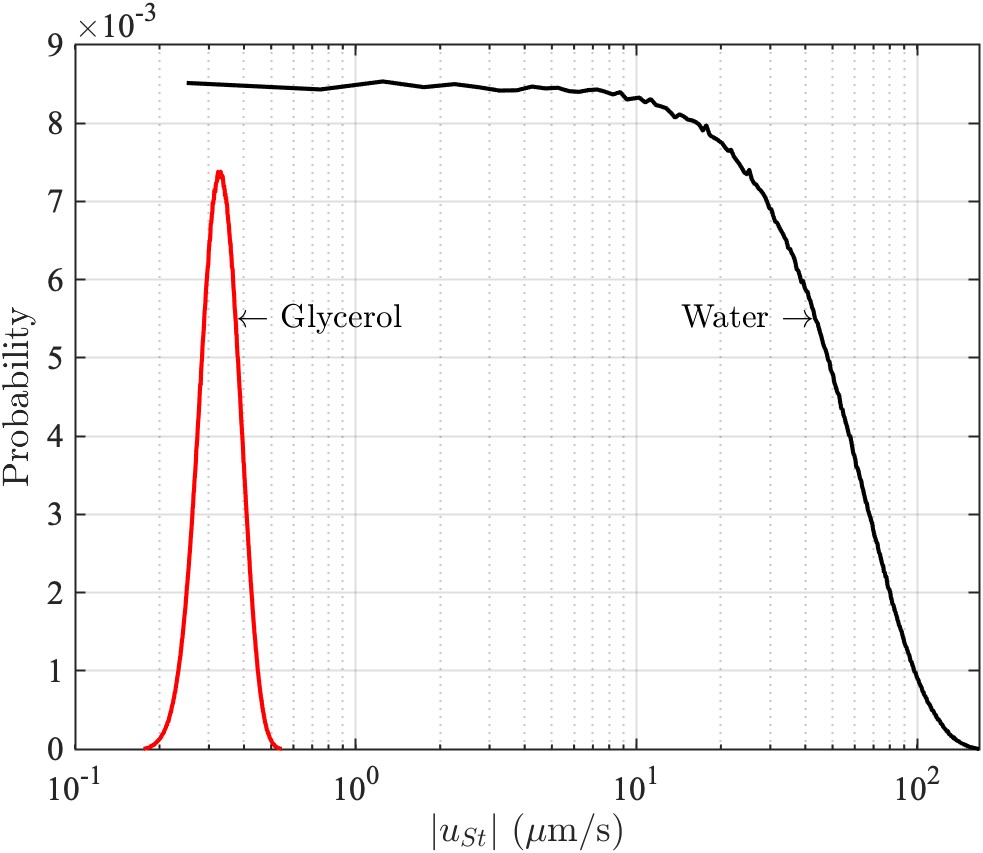}
    \caption{The probability distribution of the terminal speed magnitude for sieved, uniformly distributed, 45-53$\mu$m AGSF-33 tracer particles in the Stokes drag regime in water and pure glycerol at 23.7$\degree$C. The distribution in water is effectively symmetric about zero; only the velocity magnitude is shown here to accommodate a log-axis plot to highlight the water-glycerol distribution contrast.}
    \label{fig:agsfstokes}
\end{figure}

For a coated composite particle such as AGSF-33 the particle density $\rho_{particle}$ is given based on the coating mass fraction $\gamma$ as
\begin{equation}
    \rho_{particle} = \frac{\rho_{pyrex} (r_{pyrex}^3 - r_{air}^3) + \rho_{air} r_{air}^3 }{(1-\gamma) r_{silver}^3}
\end{equation}
where $r_i$ is the outer radius of material $i$.
To obtain an approximation of the effect of a particle-fluid density mismatch, we run Monte Carlo simulations of the Stokes settling speed (\cite{jcgm_evaluation_2008}). For simplicity, we assume that the diameter and density are independent random variables. We take the density to be normally distributed with a 0.033~g/cm$^3$ standard deviation, truncated at 0.9~g/cm$^3$ and 1.1~g/cm$^3$. We assume that Potters Beads' quality control has removed particles outside of this range. Randomly taking values from these probability distributions, we simulate equation \ref{eq:stokesDrag} $10^7$ times for particles in water and glycerol each at 23.7~$\degree$C. The resulting Stokes velocity distribution is shown in figure \ref{fig:agsfstokes}. In water, the distribution mean is $\leq O$($\pm 0.01~\mu$m/s); the standard deviation 47~$\mu$m/s. Using pure glycerol increases the fluid viscosity by an order of magnitude, dramatically reducing the particle Stokes velocity spread, as seen in figure \ref{fig:agsfstokes}. In glycerol at 23.7$\degree$C, the AGSF-33 particles rise at an average speed of 0.34~$\mu$m/s, with a standard deviation of 0.05~$\mu$m/s. It is worth noting that a tighter density distribution would be possible using sorting techniques such as floating and sinking separation. However, for simplicity, the authors used the particles as-is to best emulate an off-the-shelf product usage.

\subsection{Tungsten-Coated Hollow Carbon Spheres (CW)}
\begin{figure}
    \centering
    \includegraphics[width=\columnwidth]{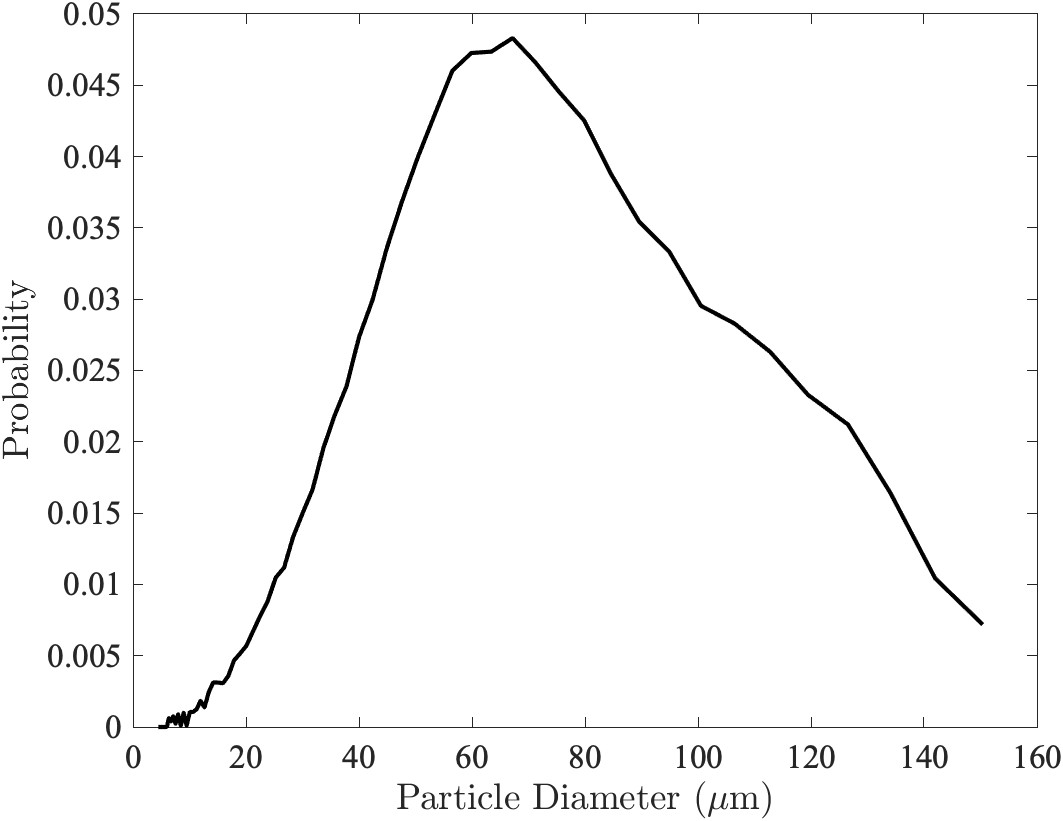}
    \caption{The CW particle diameter distribution – measured using a non-aqueous based dispersion (ISO 13320) with a Saturn DigiSizer – provided by Particle Testing Authority. While the particles are polydisperse from the factory, sieving and future manufacturing improvements can reduce the polydispersity.}
    \label{fig:cwsizedist}
\end{figure}

\begin{figure}
    \centering
    \includegraphics[width=\columnwidth]{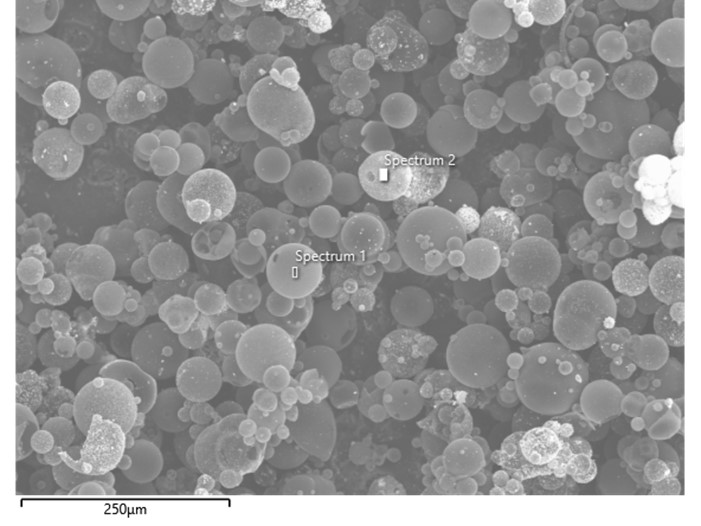}
    \caption{An SEM image of the CW tracer particles used in these experiments prior to sieving. Note the significant polydispersity prior to sieving.}
    \label{fig:cwsem1}
\end{figure}

\begin{figure}
    \centering
    \includegraphics[width=\columnwidth]{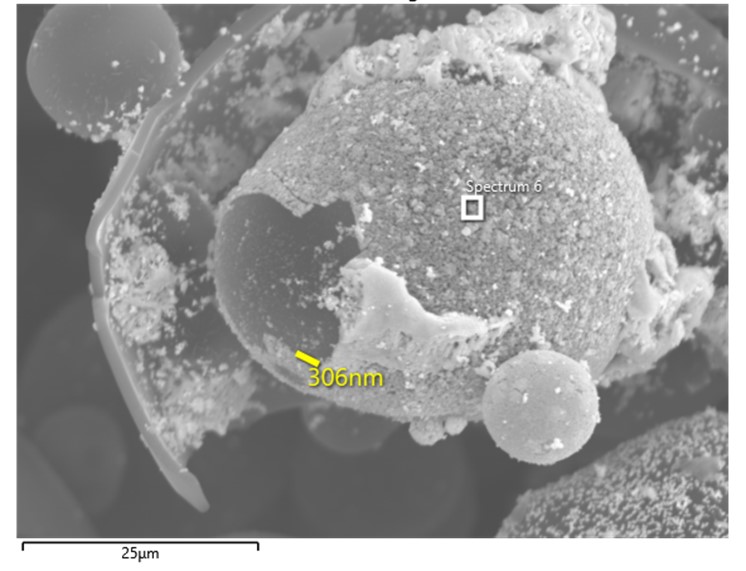}
    \caption{In this SEM image of a broken particle the coating thickness can be measured. However, we find that the coating thickness varies significantly due to the new manufacturing process not yet being fully refined. The particle shown is also non-spherical, another challenge in manufacturing new tracer particles.}
    \label{fig:cwsem2}
\end{figure}

Compared to the AGSF-33 particles, the new CW particles conceived and initially designed in the FLOW Lab and manufactured by Ultramet offer higher X-ray image contrast. A tungsten coating has a higher mass attenuation coefficient than silver, thereby generating greater contrast. In order to maintain neutral buoyancy and maximize the coating thickness, the tungsten is coated onto hollow carbon microspheres. Simulations using the Beer-Lambert method in \cite{parker_experimentally_2022} show that the CW particles can achieve a 3.4$\times$ improvement in the SNR compared to the AGSF-33 particles on a glass slide. The SNR improvement implies that CW particles will exhibit higher contrast in a fluid as well. The diameter distribution of the tungsten-coated hollow carbon spheres as-manufactured can be seen in figure \ref{fig:cwsizedist}. The polydispersity shown in figure \ref{fig:cwsizedist} is apparent in the scanning electron microscope (SEM) image shown in figure \ref{fig:cwsem1}. Figure \ref{fig:cwsem2} shows a nonspherical particle with a broken coating, demonstrating the prototype nature of these early CW particles. We take advantage of the broken coating to measure the coating thickness. Sieving reduces the polydispersity, but in general monodispersity in all properties will improve if there is wide demand and manufacturers can fine-tune their scalable manufacturing processes for making these new types of particles. In this study, the particles are sieved to the same diameter range as the AGSF-33 particles to ensure a fair contrast comparison.

\begin{figure}
    \centering
    \includegraphics[width=\columnwidth]{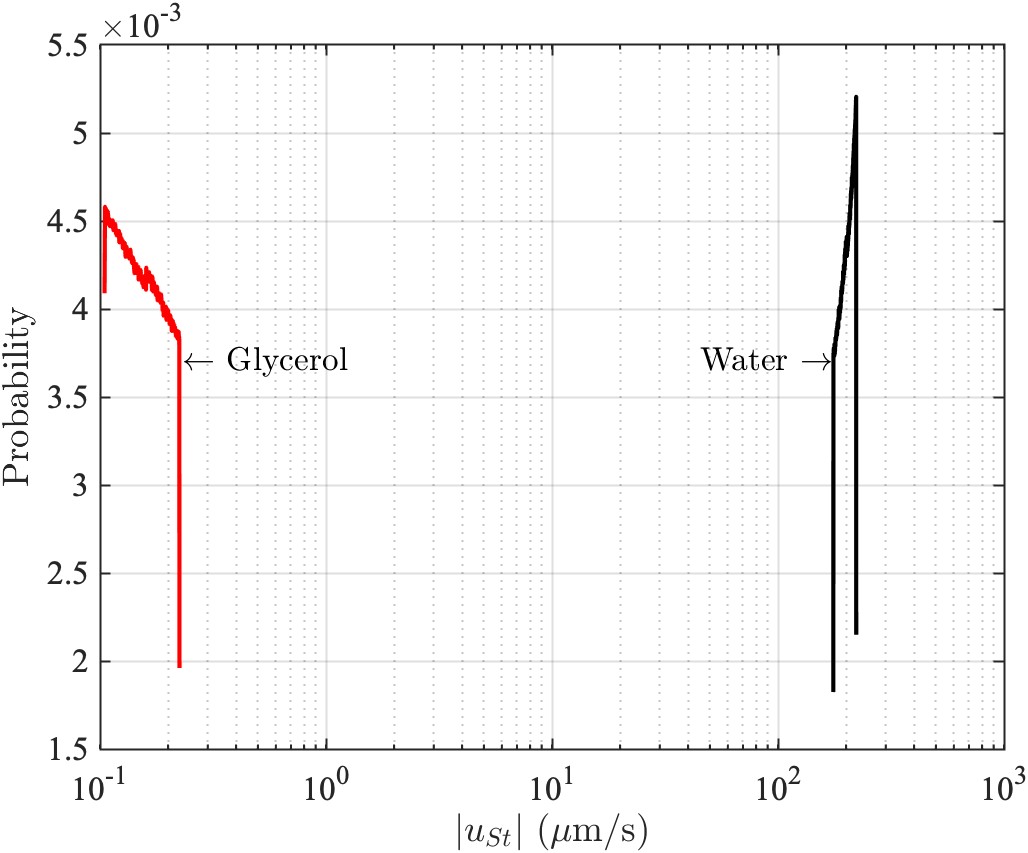}
    \caption{The Stokes terminal speed probability distribution of the CW tracer particles in water and pure glycerol.}
    \label{fig:cwstokes}
\end{figure}

We run the same settling speed Monte Carlo simulations for the CW particles as for the AGSF-33 particles in section \ref{ssec:agsf}. Based on the observed settling behavior and SEM images, the tungsten coating thickness is approximately 0.25~$\mu$m. However, in SEM images, some particles are observed to have a 0.5~$\mu$m thick coating or more, demonstrating the manufacturing variability of the CW particles in these first batches. The average hollow carbon microsphere density is $\left<\rho_{carbon}\right> = 0.5767$~g/cm$^3$. Using these data alongside the diameter distribution in figure \ref{fig:cwsizedist}, we can estimate the CW particles' Stokes velocity distribution shown in figure \ref{fig:cwstokes}. In water at 23.7$\degree$C, the CW particles settle at an average velocity of -200~$\mu$m/s with a standard deviation of 13.3~$\mu$m/s. While the mean CW particle settling speed in water is much greater than the AGSF-33 particles, in glycerol at 23.7$\degree$C, the average Stokes terminal speed of the CW particles is 0.16~$\mu$m/s, with a standard deviation of 0.035~$\mu$m/s -- a much tighter and slower distribution than in water. While these first CW particles are outperformed by AGSF-33 particles in traditional tracer particle metrics like neutral buoyancy in water and polydispersity, as more monodisperse carbon microspheres are sourced and the tungsten coating thickness is better controlled, the traditional tracer particle metrics of CW will become competitive with AGSF-33 in water as well.

\subsubsection{CW Particle Manufacturing}
The CW particles are manufactured by Ultramet using a chemical vapor deposition (CVD) process to coat carbon microspheres with tungsten. The carbon spheres are hollow, averaging a diameter of 61.74 $\pm$ 0.03~$\mu$m and a density of 0.5767 $\pm$ 0.0009 g/cm$^3$. Diameter and density are measured using a non-aqueous based dispersion (ISO 13320) using a Saturn DigiSizer and nitrogen pycnometry (ISO 12154), respectively, provided by Particle Testing Authority. The tungsten coating is applied to the microspheres with a powder bed CVD reactor. The coating thickness is varied by modification of the CVD process conditions. The coated spheres are sorted based on their diameter using various wire mesh sieves. These groups, where $d_p$ represents a particles diameter, are: $ d_p \in (0, 38]$, $(38,45]$, $(45,53]$, $(53, 63]$, and $(63, 250]~\mu$m. As mentioned in previous sections, CW particles with $d_p \in (45,53]~\mu$m are used in this study to provide the best comparison to the equivalently sized AGSF-33 particles.

\subsection{Tracer Particle Contribution to Uncertainty}\label{ssec:particleerr}

A summary of the settling speed Monte Carlo simulations for CW and AGSF-33 particles can be seen in table \ref{tab:partProp}. Ideally, the Stokes terminal velocity is at least an order of magnitude less than the characteristic velocity of the flow, i.e., $u_{St} \ll U$. When $u_{St} \ll U$, the buoyancy induced velocity can be neglected. Table \ref{tab:partProp} shows that this condition is satisfied for both particles in glycerol.

Another source of error is the particle's dynamic response to the flow. The particle response time – the Stokes time – is an indicator of the tracer particle's ability to follow a flow accurately. Assuming the relative velocity between the spherical particle and the surrounding fluid is such that the $Re(u_{rel}) \ll 1$, and the fluid acceleration is constant, the characteristic response time of the particle is given by equation \ref{eq:resptime} (\cite{m_raffel_particle_2018}).
\begin{equation}\label{eq:resptime}
    \tau_p = d_p^2 \frac{\rho_p}{18\mu}.
\end{equation}
The Stokes number, shown in equation \ref{eq:stokesNum}, is the ratio of the Stokes time to the characteristic time of the flow, $\tau_f$. For pipe flow, $\tau_f = D / U$, where $D$ is the pipe diameter.
\begin{equation}\label{eq:stokesNum}
    St \equiv \frac{\tau_p}{\tau_f} = \frac{ \rho_p d_p^2 }{ \rho_f D^2 } \frac{Re}{18}
\end{equation}
A particle is said to follow the flow accurately when $St \ll 1$. Equation \ref{eq:stokesNum} shows that smaller diameter particles are needed to measure higher Reynolds number flows, again motivating the need to explore new tracer particle designs to expand the applicability of XPTV to more flows of interest. Table \ref{tab:partProp} presents the Stokes number of AGSF-33 and CW particles in both pure glycerol and water at 23.7$\degree$C. Both particles have sufficiently small Stokes numbers for the pipe flow investigated in this study.
\begin{table}[]
    \centering
    \begin{tabular}{|c|c|c|}
        \hline
         & AGSF-33 & CW \\
         \hline
         $u_{St} / U$ (Glycerol) & 0.0013 & 0.00036 \\
         \hline
         $u_{St} / U$ (Water) & 0.017 & 382 \\ 
         \hline
         $St \times 10^{9}$ (Glycerol) & 5.2 & 6.1 \\
         \hline
         $St \times 10^{9}$ (Water) & 6.6 & 7.7 \\
         \hline
    \end{tabular}
    \caption{The Stokes number and terminal velocity ratio for the AGSF-33 and CW tracer particles for flow $Re = 0.0018$. Note that all dimensionless numbers are much less than unity in glycerol, so the particles can be expected to accurately trace the flow. AGSF-33 particles would also work as flow tracers in water at these speeds, but the prototype CW particles considered here would need to be used in flows faster than 2000~$\mu m/s$ to neglect buoyancy effects in water.}
    \label{tab:partProp}
\end{table}

\section{Results and Discussion}\label{sec:results}
\subsection{Tracer Particle Measurement Comparison}
\begin{figure}
    \centering
    \begin{subfigure}[b]{0.9\columnwidth}
        \centering
        \includegraphics[width=\columnwidth]{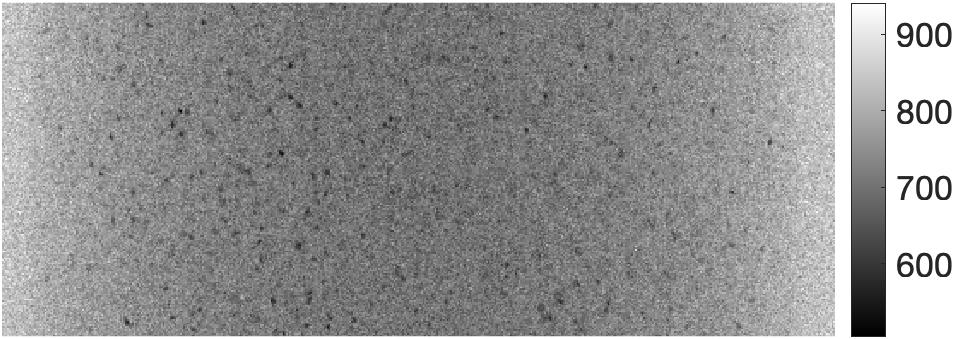}
        \caption{}
    \end{subfigure}
    \vfill
    \begin{subfigure}[b]{0.9\columnwidth}
        \centering
        \includegraphics[width=\columnwidth]{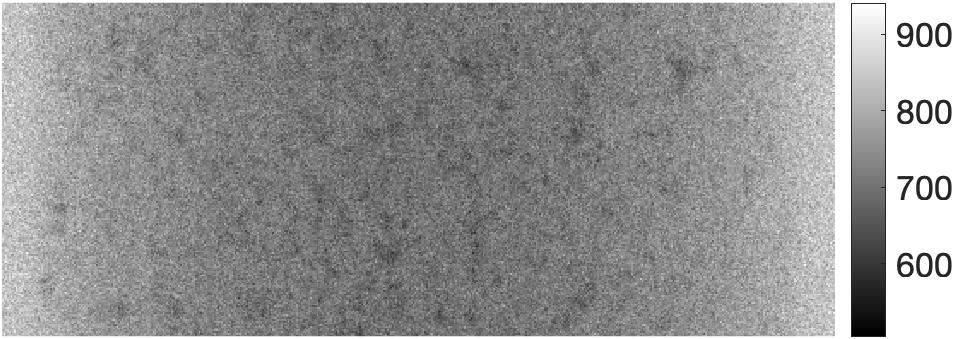}
        \caption{}
    \end{subfigure}
    \caption{a) CW tracer particles and b) AGSF tracer particles imaged with a 15~ms exposure time, 500~mm SDD, 38~mm SOD, a 55~kV acceleration voltage, and a 500~$\mu$A target current. The CW tracer particles demonstrate higher contrast than the AGSF-33 tracer particles for the same exposure time.}
    \label{fig:contrast}
\end{figure}

\begin{figure}
    \centering
    \includegraphics[width=\columnwidth]{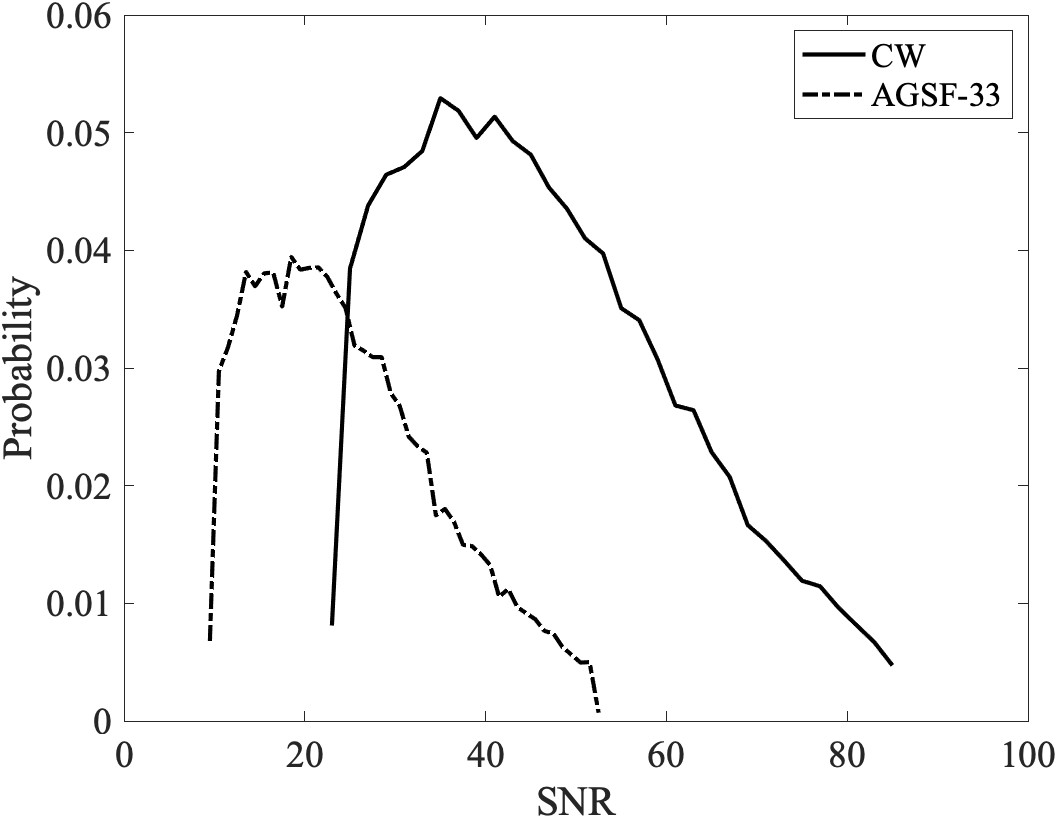}
    \caption{The measured SNR for the AGSF-33 and CW tracer particles. The CW SNR is shifted higher as a result of the higher contrast compared to the AGSF-33. Note that due to clustering of most of the AGSF-33 particles the true SNR of individual AGSF-33 particles is likely lower than observed.}
    \label{fig:snr}
\end{figure}

The benefits of the CW tracer particles stem from their marked contrast improvement. Figure \ref{fig:contrast} demonstrates the higher contrast of the CW tracer particles compared to the AGSF-33. Figure \ref{fig:snr} shows the SNR improvement of the CW tracer particles over the AGSF-33, with SNR approximated as
\begin{equation}\label{eq:snr}
    SNR \approx \frac{ \lvert \left< I \right>_{r \leq r_p} - \left< I \right>_{r_p < r \leq 3 r_p} \rvert}{ \sqrt{ \left< I \right>_{r_p < r \leq 3 r_p} } }
\end{equation}
where $I$ is the pixel intensity, $r$ is defined as the radius from the particle centroid, $r_p$ is the particle radius, and $\left< \cdot \right>$ indicates mean. The average CW particle SNR is 47, while the average AGSF-33 particle SNR is 25, for a CW to AGSF-33 SNR ratio of 1.8. Equation \ref{eq:snr} implies that if other particles are within a particle diameter of the particle edge (the definition we use for local), then the SNR calculation could be affected. Figure \ref{fig:snr} shows the general behavior of the AGSF-33 and CW tracer particle SNR, but should not be taken as a high accuracy measurement of individual particles' SNR due to the effects of dense particle seeding and 2D-projection imaging. A higher SNR improves particle localization, which yields higher precision measurements. Figure \ref{fig:contrast} also shows the AGSF-33 particles clustering, which we did not observe for the CW particles. Since the AGSF-33 particles cluster, the actual SNR improvement by CW may in fact be higher than we observed.

\begin{figure}
    \centering
    \includegraphics[width=\columnwidth]{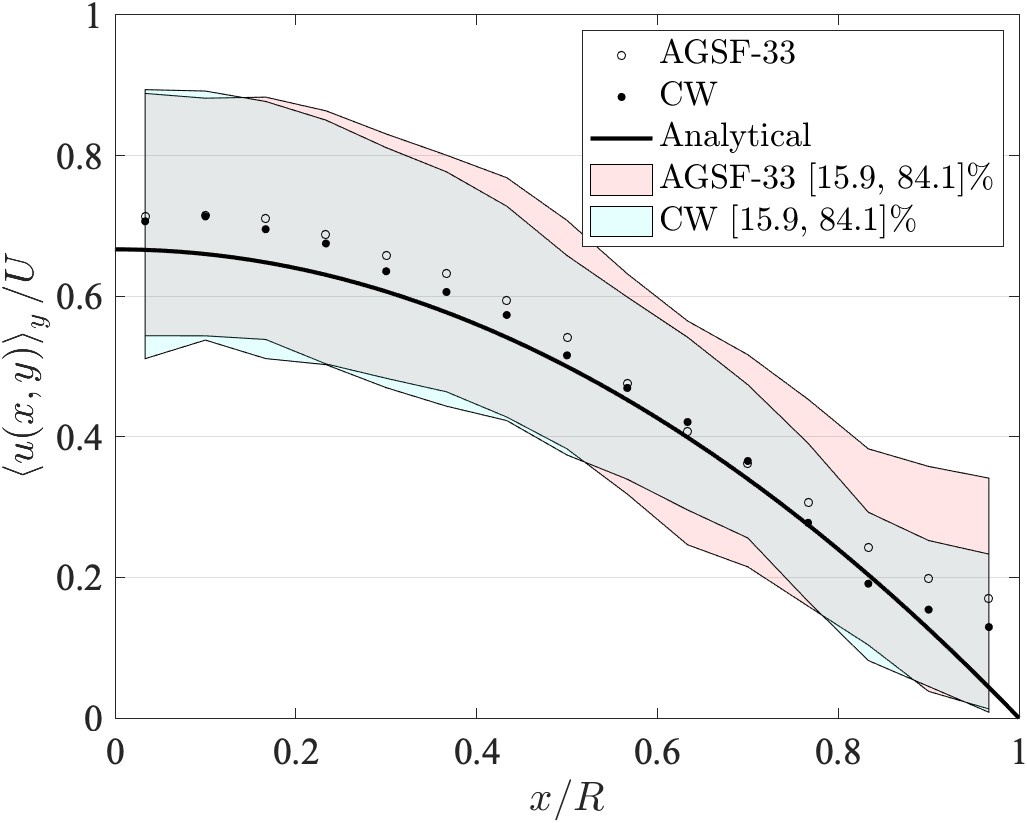}
    \caption{The DAVP as measured by the CW and AGSF-33 tracer particles compared against the analytical DAVP.}
    \label{fig:yprofiles}
\end{figure}

Figure \ref{fig:yprofiles} shows the DAVP measured with the AGSF-33 and CW tracer particles. We use the PTV algorithms in LaVision DaVis 8.4, and the details of the processing steps and settings can be found in appendix \ref{app:imgproc}. As mentioned previously, the glycerol-particle mixture is poured into the pipe at an angle, leaving room for air to escape the pipe. However, this resulted in particles being pushed to one side by shear lift. Chipped or broken particles, which are smaller, are more susceptible to these shear lift forces and are preferentially pushed to one side. These broken particles are also more buoyant than the Monte Carlo simulations in section \ref{sec:partprop} suggest. Indeed, an artifact in the velocity profile at random locations could sometimes be observed. We did not observe the same seeding distribution behavior with the AGSF-33 particles, as they are mass produced via a more refined process and broken particles have been more effectively removed. For a clear quantitative comparison less sensitive to particle distribution bias we take the average and standard deviation of both halves of the pipe combined to calculate a projection of the radial velocity profile. Both the CW and AGSF-33 particles recover the DAVP, as figure \ref{fig:yprofiles} shows. Close to the wall, the data do not go to zero because particles have a finite diameter and do not stick to the wall.

Figure \ref{fig:yprofiles} also highlights how measurement variance is exacerbated by 2D-projected XPTV. Tracer particles at the front and back walls of the pipe appear at the same $x$-location as particles at the center, so there is an unavoidably wide spread in the measured velocities. Stereo or tomographic XPTV methods are necessary to alleviate the issues with 2D-projection, and are topics of ongoing research. Light sheets like those used in PTV are not possible for XPTV. Most XPTV systems, including the one used in this study, are based on transmitted light instead of scattered light, so even if optics were available for in-lab XPTV, creating an X-ray light sheet would be not be helpful.

There is, however, an alternative to multi-source-detector pair or rotating systems for capturing 3D data. Provided that sufficiently monodisperse tracer particles are used, one can recover 3D velocity information by tracking the particle image size, which changes as a function of depth due to geometric magnification. Such a system would only require one source-detector pair. One could also filter for particle images of a certain size, effectively creating an artificial 2D imaging plane akin to a laser sheet. XPTV tracer particles with such a tight diameter tolerance are, to the best knowledge of the authors, not presently available. Such tight tolerances for the particle diameter may not be an insurmountable requirement, though.

\begin{figure}
    \includegraphics[width=\columnwidth]{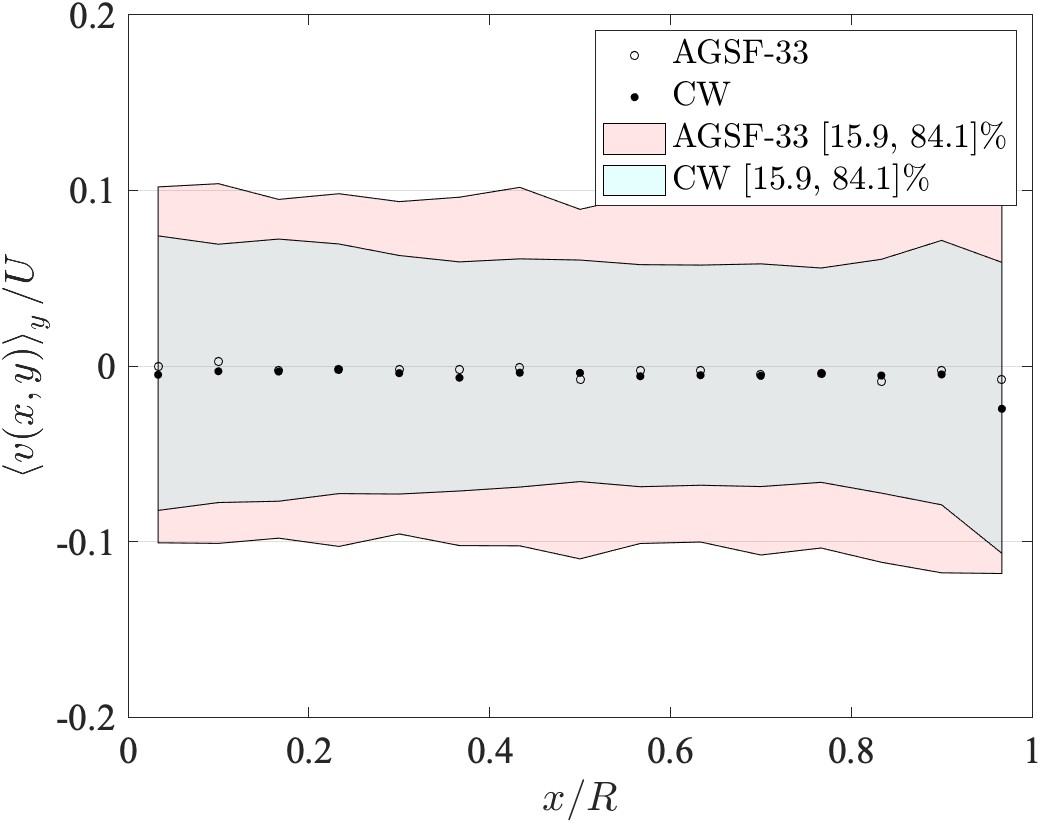}
    \caption{The measured $x$-component velocity, $v(x,y)$, for the CW and AGSF-33 tracer particles. The CW tracer particles, being more localizable and less clumped, exhibit smaller measurement variance compared to the AGSF-33.}
    \label{fig:xprofiles}
\end{figure}

Figures \ref{fig:yprofiles} and \ref{fig:xprofiles} also show that CW tracer particles are more localizable due to their higher contrast and they did not clump like the AGSF-33 tracer particles. As a result, both the axial and $x$-component velocity measurements with the CW tracer particles demonstrate less spread than the AGSF-33 tracer particles. Better localizability is crucial because it reduces measurement error in velocity and improves spatial resolution.

\begin{figure}
    \centering
    \begin{subfigure}[b]{\columnwidth}
        \includegraphics[width=\columnwidth]{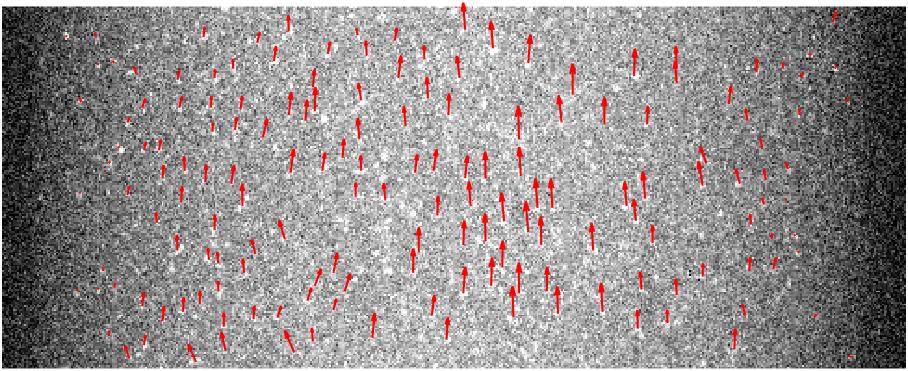}
        \caption{}
        \label{fig:cwvec}
    \end{subfigure}
    \hfill
    \begin{subfigure}[b]{\columnwidth}
        \includegraphics[width=\columnwidth]{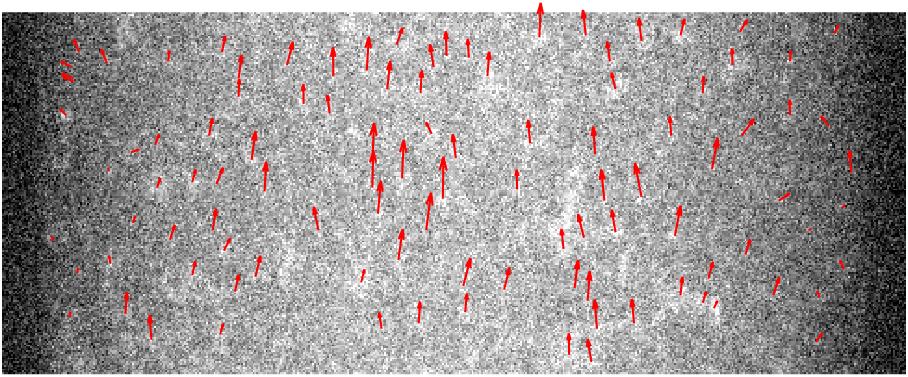}
        \caption{}
        \label{fig:agsfvec}
    \end{subfigure}
    \caption{The velocity vectors overlaid on the hundredth frame of the (a) CW and (b) AGSF-33 footage. The greater variability of the AGSF-33 $x$-component velocity due to variability of the apparent particle centroid location is apparent, whereas the CW particles offer better contrast and are more robustly localized. The color scale is inverted for PTV processing in DaVis.}
    \label{fig:vecplots}
\end{figure}

Figure \ref{fig:vecplots} shows the velocity vectors superimposed on an image. We note from inspection of many such images that: 1) the PTV algorithm does not produce spurious vectors a human observer would not create and 2) some particles that a human would trace are not traced by the algorithm. These data could likely be improved by optimizing the processing parameters or by using more advanced PTV algorithms. Improving image quality by using higher contrast particles, brighter sources, and better detectors will also help. These measurements demonstrate that time-resolved, $O(1$~cm$)$ domain XPTV at nearly 70~Hz is possible in the laboratory with both AGSF-33 and CW tracer particles. We achieve this result in spite of a dim laboratory X-ray source by using a single-threshold PCD and high-contrast tracer particles. Exploring the performance limits of CW tracer particles is the subject of future work. Brighter laboratory X-ray sources combined with advanced tracer particles will enable even faster XPTV. Particle contrast is proportional to photon flux, implying that with new X-ray sources that are 100 times brighter (\cite{hemberg_liquid-metal-jet_2003}), frame rates on the order of 1~kHz are possible.

Despite the need for future improvements in monodispersity and tungsten coating thickness control, CW tracer particles outperform AGSF-33 tracer particles in terms of image contrast and localizability while still measuring a comparable DAVP with satisfactory agreement to the analytical solution. As the tracer manufacturing process improves, CW tracer particles will enable the measurement of higher speed flows, especially when combined with brighter laboratory X-ray sources. The success and future improvement of the CW tracer particles should motivate the community and seeding particle manufactures to explore the XPTV tracer particle design space and push the boundaries of flows that can be studied with this technique.

\subsection{Benefits and Challenges of PCDs}
For the same source flux, PCDs can achieve a higher SNR and less blurring than scintillating detectors (\cite{russo_handbook_2018}). PCDs directly detect photons, dispensing with visible wavelength scatter and glow artifacts in scintillators, which both contribute to blurring. Directly detecting photons also avoids the thermal and electronic noise associated with CCD and CMOS sensors. Despite these advantages, at this stage PCDs have a number of drawbacks that require consideration.

Generally speaking, SNR increases monotonically with increasing source brightness, enabling higher frame rates. For PCDs, however, the SNR can rapidly degrade due to photon count pile-up. That is, as the number of photons arriving at a pixel per unit time exceeds the circuit's ability to distinguish individual events, there ceases to be an increasing benefit to a brighter source. Current PCD nominal count rate limits are $O(10^6$ - $10^7$)~photons/pixel/second, often with some form of retriggering, software correction, or both to mitigate the effects of pile-up. For the Dectris Pilatus3 detector used here, the nominal count rate limit is $10^7$~photons/pixel/second when using retriggering and a count rate correction. We turn retriggering and the correction off for these experiments, however, because Dectris does not recommend using them with a polychromatic X-ray source. They are calibrated at a synchrotron for use with a monochromatic source. Unfortunately, most laboratory X-ray sources are polychromatic, including the one used here. Without retriggering and the count rate correction, the Dectris Pilatus3 can handle up to about $10^6$~photons/pixel/second. Once pile-up begins to dominate, SNR degradation is observed and image artifacts such as detector tile boundaries become uncorrectable. The count rate limit of PCDs is an important limitation on the source brightness that bounds the achievable SNR.

A scintillating detector would not face the same limitations. X-rays are turned into visible wavelength photons regardless of their arrival rate, and the saturation limit for a CCD or CMOS sensor can often be adjusted. In practice, then, a scintillating detector could achieve a higher SNR than a PCD detector by enabling imaging with a brighter source than the PCD could handle. Indeed, scintillating detectors are widely used for imaging at synchrotrons, while PCDs are not placed directly in a synchrotron beamline. In practice, then, scintillating detectors can currently achieve faster frame rates or higher image quality than PCDs.

In addition to the temporal resolution advantage, scintillating detectors also have a spatial resolution advantage. Pixel sizes on currently available PCDs are relatively large – $O$(100~$\mu$m). Scintillating detectors offer much smaller pixel sizes, achieving higher spatial resolution.

Why, then, use a PCD for X-ray flow visualization? First, although PCDs currently suffer from count rate limitations, due to improving circuitry, detector materials and reduced pixel sizes, future PCDs can be expected to accommodate increasing count rates. When pile-up is not the limiting factor, the theoretical SNR advantage of PCDs is apparent. Furthermore, despite the current drawbacks, there are benefits to PCDs even now that in some cases may outweigh the limitations. Chiefly, PCDs can directly measure the deposited photon energy, in effect capturing ``color" X-ray images. PCDs that are now coming to market are able to resolve the incoming photon energy with two or more thresholds, enabling K-edge material detection. One could then track multiple tracers made of different materials, a process analogous to fluorescing tracer particles with different colors in visible light flow visualization. Additionally, one could observe the scalar mixing of two or more phases while simultaneously measuring the fluid velocity field with tracer particles, a difficult experiment with visible light techniques, and an all but impossible one in opaque media or containers.

\section{Conclusion}\label{sec:conc}
This study demonstrates quantitative laboratory XPTV by measuring the 2D-projected velocity profile of Poiseulle flow in a round pipe with two different tracer particles. By capturing images of tracer particles in 15~ms exposure times, we show the potential to speed up laboratory XPTV to nearly 70~Hz, a limit that can easily be extended with brighter sources and improved detectors. We show the value of high-contrast tungsten-coated, hollow carbon microsphere tracer particles designed specifically for quantitative XPTV. The tungsten-coated CW tracer particles offer a contrast enhancement over the previously available AGSF-33 tracer particles, making them easier to localize and, in principle, allowing the same SNR with a shorter exposure time. As the manufacturing process for the CW and other custom particles improves, their utility as XPTV tracer particles can be expected to surpass AGSF-33. Continuing exploration of the XPTV tracer particle design space is important for developing application- and technique-specific tracers. For example, to develop smaller particles or optimize the attenuation contrast for application-specific fluids. The metrics used in this study, and methods previously discussed in \cite{parker_experimentally_2022}, can be used for testing future tracer particle prototypes.

In addition to demonstrating the value of tracer particles developed specifically for X-ray flow visualization, we explored the feasibility of using PCDs for X-ray flow visualization. By showing that PCDs can acquire laboratory XPTV images at speeds comparable to scintillating detectors, we show that energy-thresholding material detection techniques are feasible for future laboratory XPTV experiments. With PCDs improving rapidly, material detection and other energy-thresholding techniques can improve tracer particle detection beyond what is capable with scintillating detectors, opening up new opportunities to use X-ray light to the experimentalist's advantage. Simultaneous multi-species and scalar field flow tracing experiments are possible with energy-resolving PCDs, for example. As laboratory X-ray sources get brighter, and PCDs get faster, these techniques will be of even greater utility. The flexibility of the system that we propose lends itself to growing alongside X-ray source and detector technology.

XPTV, XPIV, and related X-ray flow visualization techniques make possible the study of flows that opacity and refraction make difficult or impossible to observe with visible light. By demonstrating the potential of XPTV with PCDs and advanced tracer particles in the laboratory, this promising technique can proliferate in laboratories across the fluid dynamics community.

\bmhead{Acknowledgments}
We gratefully acknowledge the support of NSF EAGER award \#1922877 program managers Ron Joslin and Shahab Shojaei-Zadeh, and the additional support provided by the Society of Hellman Fellows Fund.

We also acknowledge the contributions of Angel Rodriguez for his help setting up the X-ray facility. We also  are grateful for the contributions of Andrew Kokubun who as an undergraduate researcher participated in the early stages of this project. For advice in setting up the MCNP simulations we refer to the authors also thank Prof. Massimiliano Fratoni.

\begin{appendices}

\section{Image Processing}\label{app:imgproc}
\subsection{Image Pre-Processing}
Flat-field correction is applied to all of the images. As part of this correction, dead pixels are replaced with the median value of the surrounding pixels. A flicker correction is also applied as the X-ray source intensity exhibits an $O$(0.1\%) variation at approximately 0.25~Hz and integer-multiple frequency harmonics, which is presumed to be the controller update frequency of the anode current feedback loop. The flicker correction involves determining the 1$^{st}$ and 99$^{th}$ percentile pixel values for each image. Those pixel values are then established as the saturation limits for their respective image. The images are then inverted so that the particles are bright on a dark background, which the Lavision DaVis processing algorithms expects. After inversion, the images are converted into 16-bit images. Then, the images are exported to undergo the following filtration and other operations in DaVis:
\begin{enumerate}
    \item Mask the area outside of the region of interest.
    \item Subtract the average intensity of all frames.
    \item Spatial bandpass filter to features that are 3-6 pixels in length.
    \item Subtract 3,700 counts to isolate the particle images as positive integer values in preparation for the next step. This value may change as a function of particle SNR.
    \item Set any negative pixel value to 0.
\end{enumerate}

\subsection{PIV Processing}
PIV is used as a pre-processing step to the PTV in order to establish an approximate particle velocity and a search radius for the particle tracking algorithm. The PIV settings chosen to provide a low spatial resolution estimation of the fluid velocity are as follows:
\begin{enumerate}
    \item First pass: 64$\times$64 pixel window, 50\% overlap, single pass.
    \item Second pass: 32$\times$32 pixel window, 50\% overlap, double pass.
    \item Remove groups with $<$ 5 vectors.
    \item Allowable axial vector range: 3 to 19 pixels, encompassing range physically possible given range of particle densities.
    \item Allowable x-component vector range: -4 to 4 pixels. Nominally zero, but due to shifts in the detected centroid, the $x$-component velocity can be measured as non-zero. This also allows for the $x$-axis of the images to potentially be misaligned with tube axis.
\end{enumerate}

\subsection{PTV Processing}
After the PIV processing step, DaVis' particle tracking algorithm is called. For each frame and tracked particle the centroid location, and $x$- and axial-velocity components are calculated. This data is imported to MATLAB to bin the velocities by $x$-location and calculate the average axial velocity component. The PTV settings are as follows:
\begin{enumerate}
    \item Particle size range: 3 to 12 pixels.
    \item Allowed vector range relative to reference (i.e., velocity from PIV): 10 pixels, while maximum displacement was 8 pixels.
\end{enumerate}

\end{appendices}

\bibliography{sn-bibliography}

\end{document}